\documentclass[usenatbib]{mn2e}
\usepackage[dvips]{graphicx}
\usepackage{amsmath,amsfonts,amssymb}
\usepackage{color}
\usepackage{times}
\usepackage{float}
\usepackage{hyperref}
\hypersetup{colorlinks=true,
           citecolor=blue,
           linkcolor=blue}
\definecolor{orange}{rgb}{1,0.5,0}
\newcommand{\be}{\begin{equation}}
\newcommand{\ee}{\end{equation}}
\newcommand{\bea}{\begin{eqnarray}}
\newcommand{\eea}{\end{eqnarray}}
\newcommand{\eq}[1]{\begin{equation} #1 \end{equation}}

\def\rivera{RI10 }
\def\baluev{BA11 }

\begin{document}

\title[Dynamical analysis of the Gliese-876 Laplace resonance]{Dynamical analysis of the Gliese-876 Laplace resonance}
\author[Mart\'i et al.]{\large{J. G. Mart\'i$^{1}$, C. A. Giuppone$^{1,2}$, C. Beaug\'e$^{1}$} \\
$^{1}$Universidad Nacional de C\'ordoba, Observatorio Astron\'omico, IATE, Laprida 854, X5000BGR C\'ordoba, Argentina \\
$^{2}$Departamento de F\'isica, I3N, Universidade de Aveiro, Campus de Santiago, 3810-193 Aveiro, Portugal
}

\date{}
\maketitle

\begin{abstract}

The existence of multiple planetary systems involved in mean motion conmensurabilities has increased significantly since the Kepler mission. Although most correspond to 2-planet resonances, multiple resonances have also been found. The Laplace resonance is a particular case of a three-body resonance where the period ratio between consecutive pairs is $n_1/n_2 \sim n_2/n_3 \sim 2/1$. It is not clear how this triple resonance can act in order to stabilize (or not) the systems.

The most reliable extrasolar system located in a Laplace resonance is GJ\,876 because it has two independent confirmations. However best-fit parameters were obtained without previous knowledge of resonance structure and no exploration of all the possible stable solutions for the system where done.

In the present work we explored the different configurations allowed by the Laplace resonance in the GJ\,876 system by varying the planetary parameters of the third outer planet. We find that in this case the Laplace resonance is a stabilization mechanism in itself, defined by a tiny island of regular motion surrounded by (unstable) highly chaotic orbits. Low eccentric orbits and mutual inclinations from -20 to 20 degrees are compatible with the observations. A definite range of mass ratio must be assumed to maintain orbital stability. Finally we give constrains for argument of pericenters and mean anomalies in order to assure stability for this kind of systems.
\end{abstract}


% \keywords{Celestial mechanics -- planetary systems; Planets and satellites: dynamical evolution and stability; Planets and satellites: general}

%%%%%%%%%%%%%%%%% SECCION 2 %%%%%%%%%%%%%%%%%%%
\section{Introduction}\label{introduction}
%%%%%%%%%%%%%%%%%%%%%%%%%%%%%%%%%%%%%%%%%%%%%%%
A three-body resonance is a planetary configuration were the period ratios between consecutive pairs of planets satisfies $n_{1}/n_{2} \sim p/q$ and $n_{2}/n_{3} \sim r/s$, with $p,q,r,s \in Z$. Due to the success of the \emph{Kepler} mission detecting exosystems (candidates) trapped in three-body resonances and because this type of multiresonant configurations give a possible origin to the giant planets of our own solar system \citep{Morbidelli_etal_2007}, currently three-body resonances present special interest in the scientific community.

Among the systems with three-planet resonances, \citet{Lissauer_etal_2011} found some Kepler candidates in a multi-resonant configurations (e.g, KOI-152, KOI-730 and KOI-500). The candidate KOI-152 reveals a system of three hot super-Earth candidates that are in (or near) a 4:2:1 mean motion resonance \citep{Wang_etal_2012}, having TTV signals that corresponds to gravitational interactions between them. The four candidates in the KOI-730 system satisfy the mean motion ratio 8:6:4:3. This resonant chain is a potential missing link that explains how planets that are subject to migration in a gas or planetesimal disk can avoid close encounters with each other, being brought to a very closely packed, yet stable, configuration \citep{Lissauer_etal_2011}. The KOI-500 is a (near-)resonant five-candidate system with combinations of mean motions given by $2n_2 - 5n_3 + 3n_4 \sim 1.6\times10^{-5}$ and $2n_3 - 6n_4 + 4n_5 \sim 1.3 \times10^{-5}$, maybe suggesting a strong interaction due to hidden companions in MMR \
\citep{Lissauer_etal_2011}. On the other hand, numerical studies using Saturn, Uranus and Neptune masses from \citet{Morbidelli_etal_2007} showed that is possible to obtain configurations where the period ratio between consecutive pairs 2:3 and 4:5 (or 3:4) are stable for 400 Gyr. 

The case where the period ratio between consecutive pairs is $n_1/n_2 \sim n_2/n_3 \sim 2/1$, is a particular case of a three-body resonance called {\it Laplace resonance}. This configuration is rare in our own solar system and the Galilean satellites Io, Europa and Ganymede constitute the only known example trapped in such a configuration \citep{Ferraz-Mello1979}. Among the vast variety of extrasolar planets, only GJ\,876 has been confirmed to be in a Laplace resonance. 

Unconfirmed candidates of Laplace resonance detected with radial-velocity can be found in HD40307 and HD82943. The system HD40307 has been proposed to be locked in 4:2:1 MMR and studied by \citet{Papaloizou_2010}. HD82943 was studied by \citet{Beauge_etal_2008}. The mass ratio between inner pair is $m_2/m_1 \sim 1$ and between the outer pair is $m_3/m_2 \sim 0.2$. The results were found using Stokes-like drag force with fixed values for the e-folding times for the semimajor axes and eccentricities. The authors tested under a wide variety of masses for the third body, and the planets always evolved towards a double MMR, in which $n_1/n_2 \sim 2/1$ and the ratio $n_2/n_3$ also corresponded to a ratio of integers. In a large majority of the simulations the outer pair of planets was also trapped in a 2/1 MMR (i.e. $n_2/n_3 \sim 2/1$). The dynamical analysis of the results revealed that there is an \textbf{asymmetric} libration of resonant angles. Finally a clue that would be useful in this analysis is that the 
phase space associated to the Laplace resonance is complex, and appears to be populated with a number of small islands of stable motion surrounded by large chaotic regions of instability \citep[see Fig.9 bottom in][]{Beauge_etal_2008}.  

The goal of the present work is to develop a dynamical analysis of the Laplace resonance using as primary target GJ\,876, because the system has been discussed previously in several works, giving us a well grounded basis with which to compare our results. In a parallel way we extend our understanding of the dynamical complexities of the Laplace resonance.

\begin{table*}
\centering
\caption{Four planet coplanar fits }
\begin{tabular}{ l c c c c c c c c }
\\[1ex] 
\hline\hline \\[-1.3ex]
  & \multicolumn{4}{c}{Best fit for Baluev (2011)} & \multicolumn{4}{c}{Best fit for Rivera (2010)} \\ [1ex]
\hline\\
{\bf Parameter} & {\bf Planet d} & {\bf Planet c} & {\bf Planet b} & {\bf Planet e} & {\bf Planet d} & {\bf Planet c} & {\bf Planet b} & {\bf Planet e} \\
\hline \\
   $P$ (days)                & $1.937886$   & $30.1829$  & $60.9904$  & $124.51$  & $1.937780$   & $30.0881$  & $61.1166$  & $124.26$  \\ 
   $m \, (\textrm{M}_{jup})$  & $0.0218$     & $0.747$    & $2.337$    & $0.0482$  & $0.0214$     & $0.7142$   & $2.2756$   & $0.0459$  \\
   $a$ (AU)                  & $0.02110625$ & $0.131727$ & $0.211018$ & $0.33961$ & $0.02080665$ & $0.129590$ & $0.208317$ & $0.3343$  \\ 
   $e$                       & $0.178$      & $0.2498$   & $0.0328$   & $0.008$   & $0.207$      & $0.25591$  & $0.0324$   & $0.055$   \\ 
   $\omega \, (^{\circ})$      & $224.0$      & $252.08$   & $248.7$    & $181.0$   & $234.0$      & $48.76$    & $50.3$     & $239.0$   \\
   $\psi \, (^{\circ})$        & $357.6$      & $71.09$    & $341.13$   & $299.3$   & $229.0$      & $343.35$   & $16.0$     & $234.0$   \\
   $M \, (^{\circ})$          & $133.6$      & $179.01$   & $92.43$    & $118.3$   & $355.0$      & $294.59$   & $325.7$    & $335.0$   \\
\hline
\end{tabular}\label{Table1}
\\[1ex]
\begin{flushleft}
Four planet coplanar fit for GJ\,876 with $i = 56.1^{\circ}$ from \citet{Baluev_2011} and with $i = 59.^{\circ}$ for \citet{Rivera_etal_2010}. The masses listed in the table are corrected by the corresponding value of the inclination for each fit. The coordinates are given in Jacobi reference frame. In his work, Baluev gives the angle $ \psi = M + \omega$, so, the calculated value of $M$ is also written (in the same way for Baluev).
\end{flushleft}
\end{table*}

\section{GJ\,876}\label{sec2}

The GJ\,876 system contains four confirmed planets, and as an exoplanetary system it is really unique. GJ\,876 was the first system detected locked in mean-motion resonance. The planets GJ\,876 b and c, have mass ratio $m_c/m_b\sim$ 3 and orbital periods of 61.11 and 30.08 days respectively \citep{Rivera_etal_2010} and their strong interaction is evident in radial velocity because their small semimajor axis ($<0.2 AU$) and small primary star ($M_{\bigstar}$ = 0.32 for \rivera and $M_{\bigstar}$ = 0.334 $M_\odot$ for \baluev). An extensive bibliography on this system brought major contributions to the development of models and methods, including: detection using mutual interaction \citep{Rivera_Lissauer_2000, Laughlin_Chambers_2001}, planetary migration and resonance capture \citep{Kley_etal_2005}, and periodic motion in planetary resonances for massive planets \citep{Hadjidemetriou2002, Beauge_Michtchenko_2003}.

From the very beginning, the possibility of hosting additional planetary bodies have been subject of intense studies and the last and outermost planet (GJ\,876 e) has been confirmed by two independent works \citep{Rivera_etal_2010, Baluev_2011} to be in a 3-body resonant configuration (e.g. Laplace resonance) with planets GJ\,876 c and b. A detailed stability analysis for the system containing an additional 15 days period planet discarded its presence \citep{Correia_etal_2010, GerlachHaghighipour_2012}. 

\subsection{Resonant variables}
Consider a system with three planets in the Laplace configuration. Then, the resonant variables involved in this problem are the resonant angles of the two two-body resonances taken separately

\eq{
\begin{split}
  &\theta_{1} = \lambda_{1} - 2\lambda_{2} + \varpi_{1}\\
  &\theta_{2} = \lambda_{1} - 2\lambda_{2} + \varpi_{2}\\
  &\theta_{3} = \lambda_{2} - 2\lambda_{3} + \varpi_{2}\\
  &\theta_{4} = \lambda_{2} - 2\lambda_{3} + \varpi_{3}.
 \end{split}
\label{30}}

\noindent where $\lambda$ is the mean longitude and $\varpi$ is the argument of pericenter. We reserve the subscripts 1 to 3 for orbital elements and masses from innermost to outermost planet involved in the Laplace Resonance.

We can also define the Laplace variable (or $\theta_{5}$) which can be written as a combination of two of the resonant angles:

\eq{
\theta_{5} = \theta_{2} - \theta_{3} = \lambda_{1} - 3\lambda_{2} + 2\lambda_{3}.
\label{31}}

\noindent being the secular variables:
\eq{
\begin{split}
  &\Delta\varpi_{21} = \varpi_{2} - \varpi_{1} = \theta_{2} - \theta_{1}\\
  &\Delta\varpi_{32} = \varpi_{3} - \varpi_{2} = \theta_{4} - \theta_{3}
 \end{split}
\label{32}}

The libration/circulation regime and libration amplitudes of these angles give much information about the resonant dynamics. It is clear from equation \eqref{31} that a simultaneous libration of the angles $\theta_{2}$ and $\theta_{3}$ means that the Laplace angle is also librating.

Simplified dynamical models for this type of resonant configurations can be found in \citet{LithwickWu2012} and \citet{BatyginMorbidelli2012}. These models were obtained as a way of introducing dissipative effects in the evolution of resonant configurations, and are too simplified for a general dynamical analysis of three-body resonances. On the other hand, numerical simulations of capture into the Laplace resonance due to gas-driven migration were performed in \citet{Libert_Tsiganis_2011}, for different mass ratios and drag parameters. Results show that if one of the inner planets has its eccentricity pumped to a value higher than 0.3, then the mutual inclination can be excited up to 30$^{\circ}$.

We use here the best fit model for GJ\,876 as a first step trying to understand this Laplace-resonance configuration.

%%%%%%%%%%%%%%%%% SECCION 2 %%%%%%%%%%%%%%%%%%%
\subsection{Dynamics for best fits}\label{stability}
%%%%%%%%%%%%%%%%%%%%%%%%%%%%%%%%%%%%%%%%%%%%%%%
\begin{figure}  
\centering
\mbox{\includegraphics[width=9.0cm]{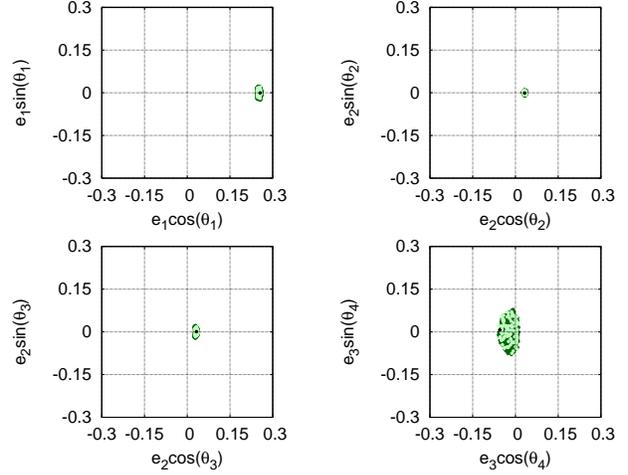}}
\caption{Time evolution of the resonant angles $\theta_{i}$, $i = 1, 2, 3, 4$ for our integration with initial conditions taken from Rivera's fit.}
\label{fig2}
\end{figure}

Table \ref{Table1} gives the best fits planetary masses and orbital parameters from \citet{Rivera_etal_2010} and \citet{Baluev_2011} (\rivera and \baluev respectively from now on). The inner pair has a inner planet significantly less massive than the outer (e.g. GJ\,876 $m_b / m_c \sim 3$), meanwhile for the outer pair the mass ratio is $m_e / m_b \sim 0.02$. We integrated initial conditions from Table \ref{Table1} and obtained qualitatively the same results than \rivera and \baluev.

The libration centers (around zero) and amplitudes are similar for the angles $\theta_{1}$, $\theta_{2}$ and $\theta_{3}$, while the last of the resonant angles $\theta_{4}$ is librating around $180^{\circ}$ with the same amplitude for both published solutions. This last result differs slightly from results given in \rivera where they claim that the $\theta_{4}$ is circulating. Some values for $\theta_{4} \sim 0^{\circ}$ are due to the small eccentricity of the planet, thus apparently giving circulation instead of libration around $\theta_{4} = 180^{\circ}$.
 
We plotted in the ($e_{j}\sin\theta_{i}, e_{j}\cos\theta_{i}$) space the initial conditions from \rivera in Figure \ref{fig2}. We used color code for different total integration times: $10^{2}$ years (light-green), $10^{3}$ (dark-green), $10^{4}$ (light-blue) and $10^{5}$ years (dark-blue). The blue dots cannot be seen because no change in the libration region is evident after the first $10^{3}$. We checked that the secular angles $\Delta \omega_{32}$ and $\Delta \omega_{31}$ librate around $180^\circ$, while $\Delta \omega_{21}$ librates with a very low amplitude around zero (being the same behavior found by \rivera and \baluev).

The Laplace angle $\theta_{5}$ is librating around zero with an amplitude $\sim 40^{\circ}$ (i.e. $\theta_{5} = 0 \pm 40^{\circ}$), for at least $3 \times 10^{6}$ years in both published fits. This amplitude is also consistent with the value given in \rivera. The non-regular libration of the Laplace angle is evident in the first 100 years of integrations.

We integrated the two solutions listed in Table \ref{Table1} including calculations of various chaoticity indicators which identified this systems as irregular. This chaotic nature of the systems is later discussed and analyzed in more detail in sections \ref{sec3} and \ref{sec5}. Although chaotic, both systems are long-term stable, and it is worth to analyze them in a thorough way.

Rivera and Baluev show that their solutions are stable for at least $10^{6}$ years. These long term integrations contain thousands of secular periods of the system, corresponding to roughly $10^{8}$ to $10^{9}$ years when rescaled to the subsystem of the giant planets in our Solar System. Baluev's long term integrations were done with initial conditions slightly different from those listed in Table \ref{Table1}. The parameters were determined with the fitting model that accounts the white and red noise and fixing the eccentricity of the outermost planet at $e_{e} = 0$. Also long-term integrations were done with only three planets, not taking into account the innermost.

%Although chaotic, the systems appear stable and any instability will not emerge for at least $3 \times 10^{6}$ years.

\begin{figure}[h]
\centering
{\includegraphics*[width=\columnwidth]{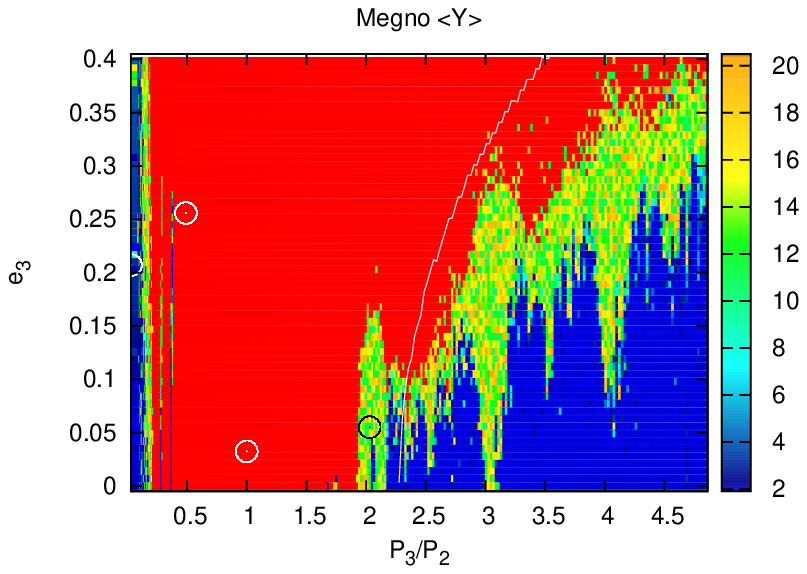}}\\
{\includegraphics*[width=\columnwidth]{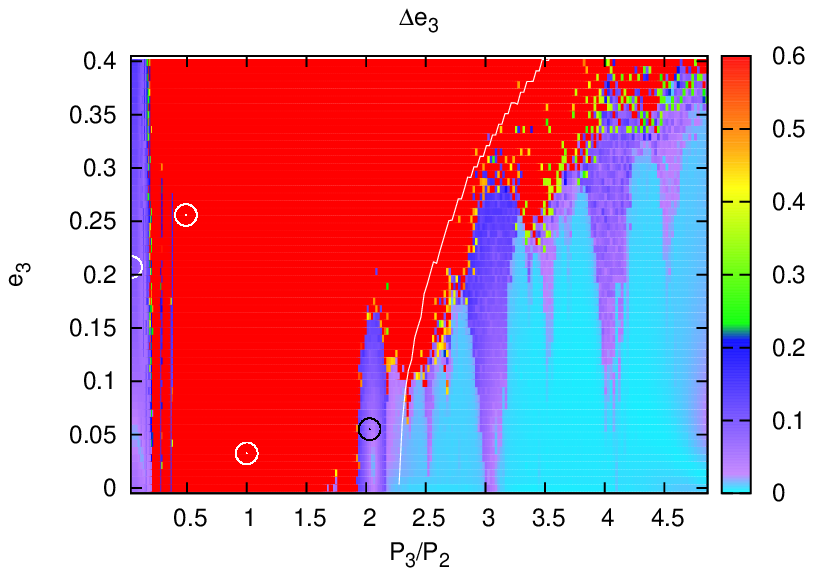}}\\
{\includegraphics[width=\columnwidth]{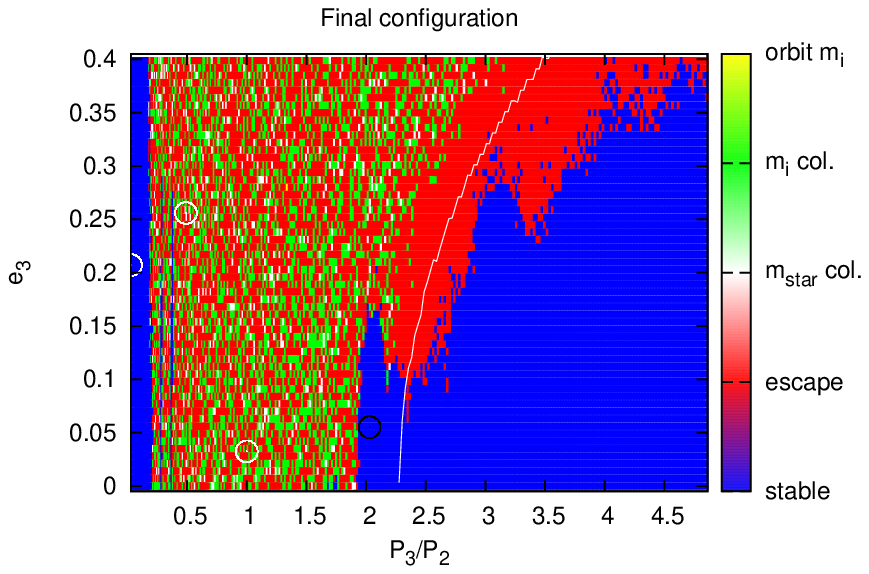}}
\caption{Stability maps for an exterior planet in Laplace resonance. {\bf Top panel}. Megno chaos indicator. {\bf Middle panel} $\Delta e_3$. {\bf Bottom panel}. Test planet end-states (stable, escape from system ($a_3> 5$ AU), collision with central star $M_{\bigstar}$ ($a_3< 0.05$ AU), collision with another companion (when mutual distance with planet $i$ is lower than sum of their mutual radii), and capture orbit around $m_i$. The white line was calculated using \citep{Gladman_1993}. White circles are the locations in the plane for planets $1$ and $2$. The phase space of the system is explored by varying the semi-major axis $a$ and eccentricity $e$ of the outer planet marked as black circle.}
\label{fig3}
\end{figure}

\subsection{Dynamics around best fits}\label{sec3}

Throughout the numerical integrations we choose the mass of the central star accordingly to each fit, in the case of \rivera $M_{\star} = 0.32 M_{\odot}$ while in \baluev $M_{\star} = 0.334 M_{\odot}$. The integrations are for coplanar systems. However we checked some runs with nearly coplanar systems and the general results remained almost unaltered. 

As mentioned in the previous section, the best fits resulted chaotic although long term stable. This led to the question about the behavior of the system surrounding the best fits. The dynamics of the Laplace resonance present in the GJ\,876 is studied in this work through several stability maps carried out with a Burlisch-Stoer based N-body code (precision better than $10^{-12}$) using as initial conditions Jacobi osculating variables.

The numerical integrator calculates the MEGNO value (Mean Exponential Growth of Nearby Orbits) for each initial condition. For a detailed explanation of MEGNO ($<Y>$) we refer the reader to \citet{Cincotta_2000}, \citet{Cincotta_Giordano_Simo_2003} and \citet{Maffione_etal_2011}. This indicator has the great feature of identifying chaotic orbits in less CPU time than other indicators (e. g. lyapunov characteristic exponent, etc), however it cannot give a precise representation of the structure of a resonance, as it only differentiate regular ($<Y> \sim 2$) from chaotic orbits ($<Y> \gg 2$). 

For each initial condition we also calculated the maximum amplitude of variation of eccentricity attained during the integration ($\Delta e_{i}$ for each planet). Due to angular momentum conservation the lower mass planet exhibits higher variation ($\Delta e_{3} > \Delta e_{2}, \Delta e_{1}$). This indicator discriminates different kind of dynamics of the system. The most drastic changes in the eccentricity of a planet is when the body is located near the separatrix of a resonance, although the amplitude of variation has an upper bound. Thus $\Delta e_{3}$ can be used to define precise limits to resonance widths and extensions, without expensive additional numerical calculations. 

We used these two indicators because they are complementary and they provide different kind of information.

\subsection{Chaoticity  Maps}

We performed a series of maps around Rivera's solution from Table \ref{Table1} to find clues for the observed stability/chaos in the GJ\,876 system. We first integrated a grid of initial conditions for the outermost planet in the $(a_3,e_3)$ plane, taking its mass equal to $14.6 M_\oplus$. Figure \ref{fig3} shows results in the region $a \in (0.03,4.86)$ and $e \in (0.0,0.4)$. The colors show the MEGNO (top) and $\Delta e_{3}$ (middle) values of each initial condition integrated for $5\times10^4$ years. We identified the location of planets 1 and 2 with white circles and the nominal location of planet 3 with a black circle. Red color-codes were reserved for orbits that disrupt within the total integration time.

Probably the first insight about the stability of the system can be given by testing if the system is stable according to the stability criteria developed by \citet{MarchalySaari_1975}. This criteria uses energy and angular momentum of a three-body system to estimate the Hill-stability limit. The problem of trying to fit this limit to our system is that we have essentially a four-body system. It is also a criteria developed considering the secular evolution of the system, which is not the case for a system trapped in MMR's. Even with these reservations in mind, we have plotted the analytical stability limit as a white curve. The agreement with the N-body simulations is not clear. % In other words the system cannot be in a MMR that could be protecting the system from close encounters. This limit is shown in Figure \ref{fig3} as a white curve, and is clearly not a good choice for analyzing stability of resonant systems such as GJ\,876.

Although these maps were done using Rivera's best fit, the same maps were also carried out around Baluev's best fit. The results are almost independent of the fit used for the GJ\,876 system, probably evidencing the robustness of this approach.

MEGNO color-code identified the location of Rivera's best fit inside a chaotic island (Fig. \ref{fig3}-top). Although not surprising, because the best fit configuration has a chaotic value of MEGNO, this island appears to be surrounded by highly unstable regions. The x-axis was rescaled in order to show the period ratio $P_3/P_2$, and easily indentificate the vertical spikes at the true location of several MMR; namely from left to right 2:1, 5:2, 3:1, 7:2, 4:1, and 9:2. MEGNO identifies all the resonances as chaotic. When we plot the same grid using a color-code for the variation of eccentricity (Fig \ref{fig3}-middle) we can see more clearly the MMR structure. {Outside the MMRs the eccentricity variation of the exterior planet remains almost unaltered ($\Delta e_3 \sim 0$) while inside the MMRs variations of the order of $\Delta e_3 \lesssim 0.2$ are observed. When two isolated resonances overlap each other the system becomes unstable, explaining the origin of red region below the \citet{MarchalySaari_1975}
 limit.}
Fig \ref{fig3}-bottom shows the test planet final states (stable, escape, collision or capture). All the unstable conditions in red located to the right of the white line calculated with \citet{MarchalySaari_1975} criteria correspond to orbits that escape from the system, thus Hill-stable configurations. Initial conditions to the left of this line can have several final states, however capture around the other planets seems to be improbable, maybe due to the strong interactions between the inner pair of planets involved in Laplace resonance. 

\begin{figure}
\centering
{\includegraphics[width=\columnwidth]{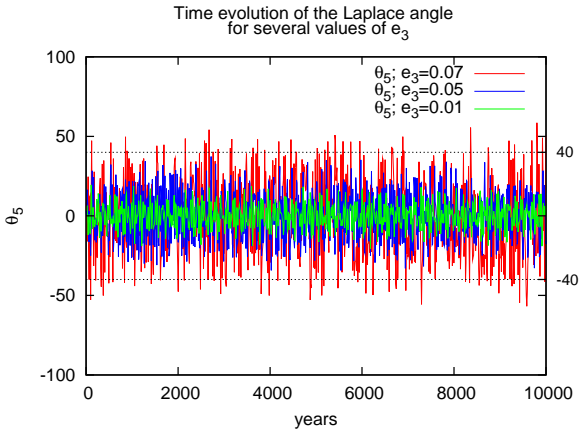}}
\caption{Evolution of Laplace angle depending on initial eccentricity for the exterior planet. It seems that $e_3\sim 0.015$ gives the minimum amplitud for the oscillation.}
\label{fig3b}
\end{figure}

In the top frame of Figure \ref{fig3} we can see a strip located at almost the nominal semimajor axis but for eccentricity around $e_3\sim0.03$ where MEGNO values are equal to 2 (regular orbits) and $\Delta e_3 \sim 0$. Thus we identify this strip as the probable center of the resonance for this system. We integrated several conditions in this strip to show how the amplitude of Laplace angle varies. Figure \ref{fig3b} shows the variation of $\theta_{5}$ with time for three of these runs. It clearly shows the tendency of the Laplace angle's libration amplitude to go to zero for values around $e_{3} = 0.015$. 

As stated before, although the best fits of Table \ref{Table1} are cataloged as chaotic, this first sight at the phase space surrounding the initial condition for the best fit can give us some clues on the long-term stability of the Laplace resonance of GJ\,876. In fact, the location of the outermost planet best fit relative to this regularity strip at the 2:1 MMR with planet 2 is probably causing the system to be long-term stable despite its chaotic nature.

Three-body resonances are much more complex than their two-body counterparts. We have more free parameters to sample in our maps. Thus, we proceeded to analyze other planes to gain further insight.

Because the pair eccentricity and argument of perihelion ($e_3,\omega_3$) are highly correlated and not very well constrained from radial velocity determinations \citep[see e.g.][]{Giuppone_etal_2009} it is expected that the GJ\,876 system has a lack of precision in those two parameters. Thus we performed a search within the plane ($e_3,\theta_4$) to see if we could find a regular solution throughout the resonant angle $\theta_4$=$\theta_4$($\omega_3$). 

We performed these ($e_3,\theta_{4}$)  maps for Rivera's and Baluev's fits, and computed the values of $<Y>$, and $\Delta e_{i}$, $i = 1, 2, 3$ as well as disruption times of the original system. The stable island where the best-fit from Table \ref{Table1} is located is surrounded by highly unstable region (not surviving after $\sim 10^3$ years of integration) in red color. 

\begin{figure}  
\centering
{\includegraphics*[width=\columnwidth]{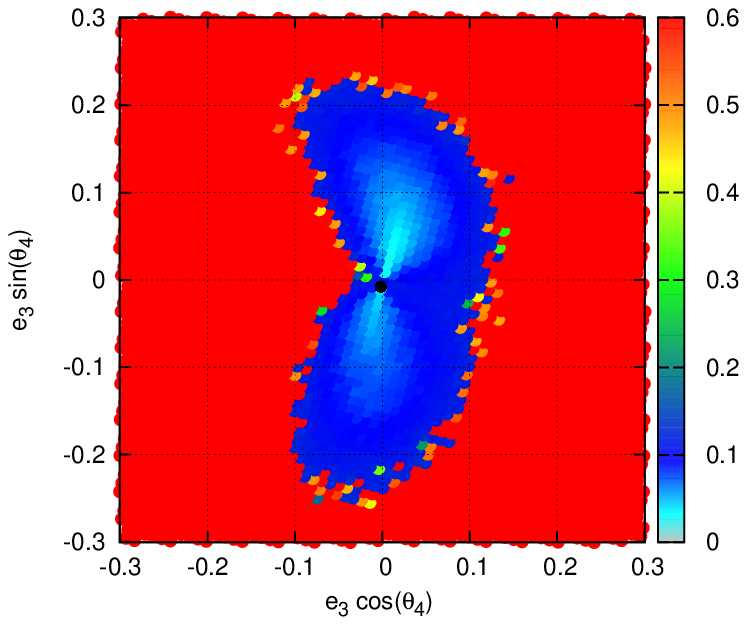}}\\
{\includegraphics*[width=\columnwidth]{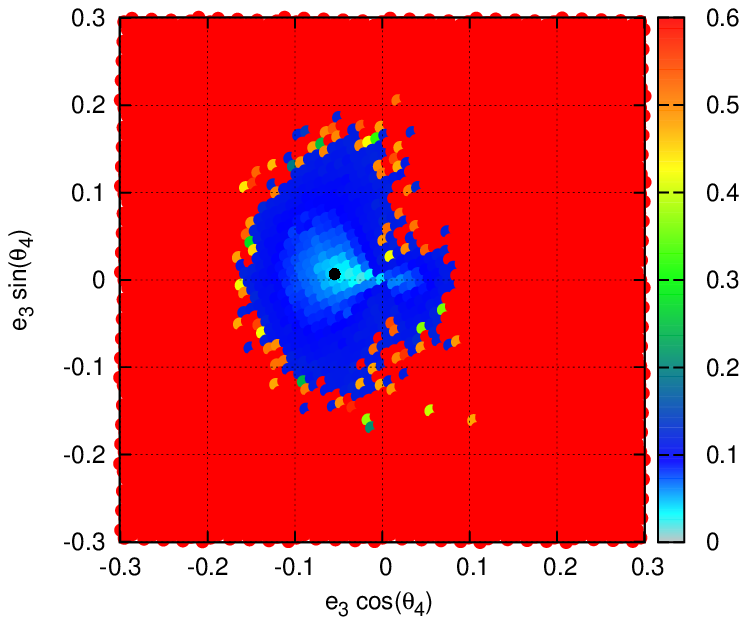}}
%\begin{flushleft}
\caption{$\Delta e_{3}$ indicator for exterior planet in the plane  ($e_3 \cos(\theta_4),e_3 \sin(\theta_4)$). Top (Bottom) using Baluev (Rivera) solution for the GJ\,876 system. The black dot represents the configuration from fits.}
%\end{flushleft}
\label{fig5}
\end{figure}

The results are shown in Figure \ref{fig5} where only the value of $\Delta e_{3}$ is shown in colors (the Megno values for this maps shows that the region sampled does not exhibit regular dynamics). Despite this, we can see in Figure \ref{fig5} that the $\Delta e_{3}$ is a much better dynamical indicator. A structure of the space can clearly be distinguished where, judging by the values for the total disruption times, the dynamics is stable at least for $10^{5}$ years. This island is present at both plots, top and bottom, and they seem to differ in their orientation. Depending on the best fit, the orientation of the stable island seems to be oriented ($0^\circ,180^\circ$) Rivera (Figure \ref{fig5}-Top) or ($90^\circ,270^\circ$) Baluev (Figure \ref{fig5}-Bottom). At the center of this stable structure we identify that the orientation of the region with $\Delta e_{3} \sim 0$ is similar to the outermost stable structure.

Finally we must address that the same results were obtained without considering the innermost planet (those with period $P\sim 1.93 d$), thus there is no need to consider its presence as a chaoticity enhancer of the system. Due to this similarity of results without planet d, we continued to work with only the three outermost planets of the GJ\,876 system.

\section{Limits to the stability}\label{sec5}

In a more general view, we tested some stability (although not regularity) limits for a multi-planetary system locked in a Laplace resonance. We performed a series of grids varying the mass, inclination and eccentricity of the the outermost planet of the system GJ\,876. We used the same chaoticity indicators as in previous section. Although not significantly different, the grid evaluating the $\Delta e_{3}$ is the clearest because it shows very sharp limits for the long-term stability of the system. 

\begin{figure}  
  \centering
  {\includegraphics*[width=\columnwidth]{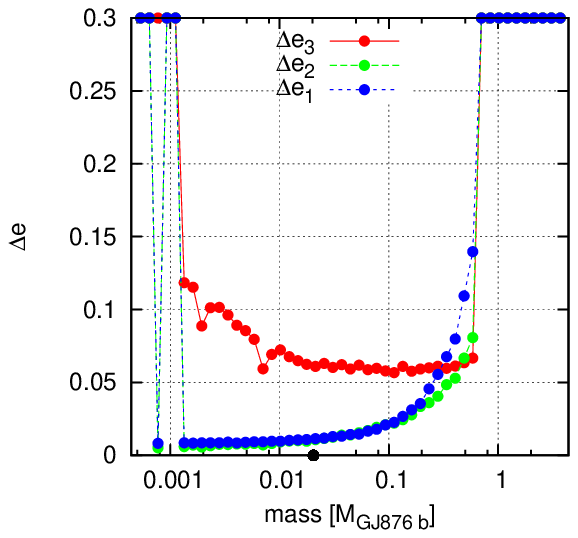}}\\
  {\includegraphics*[width=\columnwidth]{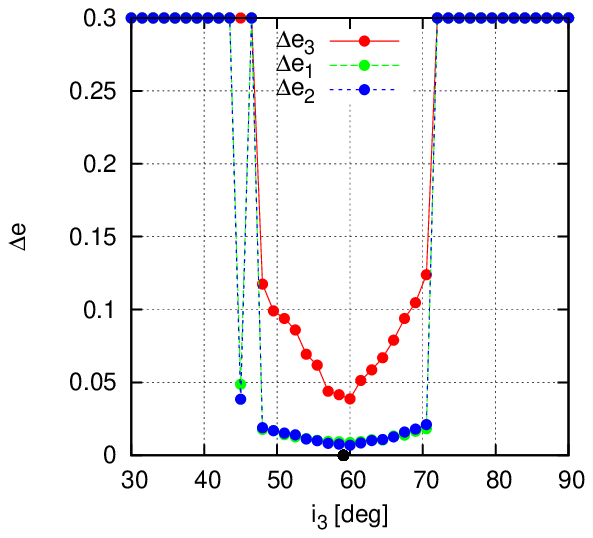}}
    \caption {Top:$\Delta e_{i}$ indicator for different masses of the exterior planet in units of intermediate companion using  Rivera's solution for the GJ\,876 system. The black dot represents the configuration from fits. Bottom: $\Delta e_{i}$ indicator for different inclination of the exterior planet. Rivera's or Baluev's solution initial conditions give qualitatively same results. The black dot represents $59^o$.}
  \label{fig6}
\end{figure}

First, we performed integrations of the GJ\,876 configuration from Rivera's and Baluev's best fit changing the mass of the exterior planet from $10^{-3}$ to $10 M_{jup}$ (i.e. corresponding to $m_3/m_2 \in (4 \times 10^{-4}$ to $4.2$). Figure \ref{fig6}-top shows that there is a mass range were we can have stable solutions trapped in a three body Laplace configuration. There are not stable systems in a Laplace resonance with an infinitesimal outermost planet. As the mass of the exterior planet increases, the stronger interaction with the interior bodies causes the innermost planet eccentricity to vary up to 0.1. After $m_3/m_2 > 0.5$ (i.e masses greater than 1.27 $M_{jup}$) the system no longer survives the $10^{6}$ years. We can see that the variation of the eccentricity of the exterior planet is compensated with the variation of the interior ones as its mass $m_{3}$ gets bigger, and the perturbation increases.

Taking again initial parameters from Table \ref{Table1} we varied only the inclination of exterior body with the nodal longitude for all the planets set to $0^{\circ}$ and analyzed the chaos/stability. Although chaotic, the system is stable for $10^6$ years for inclination $i_3$ of exterior planet from $46^o$ to $72^o$, corresponding to mutual inclinations from $-15^o$ to $15^o$. Outside this range the system quickly collides (Figure \ref{fig6}-bottom). The minimum variation of eccentricity for the less massive planet is produced when the system is coplanar, thus the most probable configuration. This results are also in agreement with \cite{Libert_Tsiganis_2011}, where they show that when eccentricities of the planets remain below $\sim 0.35$, and excitation of the inclinations does not occur.

\begin{figure}  	
  \centering
  {\includegraphics*[width=\columnwidth]{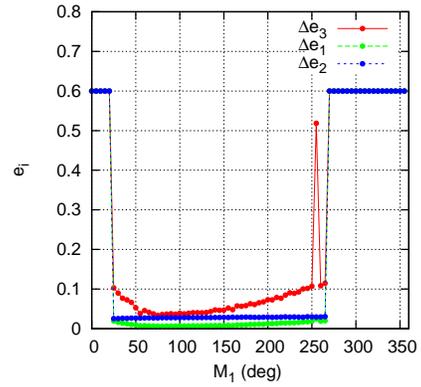}}
    \caption{$\Delta e_{i}$ indicator for different values of the mean anomaly of the exterior planet. These conditions were integrated for $1\times10^6$ years, and it seems that the value of $M_{1} = 60^{\circ}$ is the most regular condition.} 
  \label{fig7}
\end{figure} 

To gain independence of initial values from previous works, we found a representative plane were all angles were set to $0^\circ$ ($\theta_{1}=\theta_{2}=\theta_{3}=\Delta\varpi_{31}=0$) but $\Delta\varpi_{32}$=$180^\circ$ (according to the results of long term integrations). The only parameters with free values are the mean anomalies, in fact related between them: $M_1=2M_2=4M_3$.

We integrated the systems using this representative plane, varying $M_1$ from $0$ to $360^\circ$ and using all other parameters as given in Table \ref{Table1}. Figure \ref{fig7} shows the variations of eccentricities $\Delta e_{i}$ for the three planets depending on $M_{1}$, integrated for $10^6$ years. The other indicators showed same behavior. The aim of this $M_{1}$-grid was to identify the possible value for $M_{1}$ where the system appears less chaotic in order to have a set of values for the angular parameters which are more suitable for analyzing the stability maps of the system. All the systems are chaotic, but we identified very sharp limits to the long-term stability of the system: values for $M_{1}$ from $20^{\circ}$ to $260^{\circ}$ are long-term stable for at least the total integration time-span.
Because the system with $M_{1}=60^{\circ}$ exhibits the lower amplitude of variation of the eccentricity $e_{3}$, we accordingly choose the initial values of $M_{2}=120^{\circ}$ and $M_{3}=240^{\circ}$ and ran integrations in a $(a_{3},e_{3})$ grid. We recall that unstable solutions for $M_1 \in [0^{\circ}, 20^{\circ}]$ as well as $M_1 \in [260^\circ,360^\circ]$ are unstable in less than $10^5$ years.

\begin{figure}  
{\includegraphics*[width=\columnwidth]{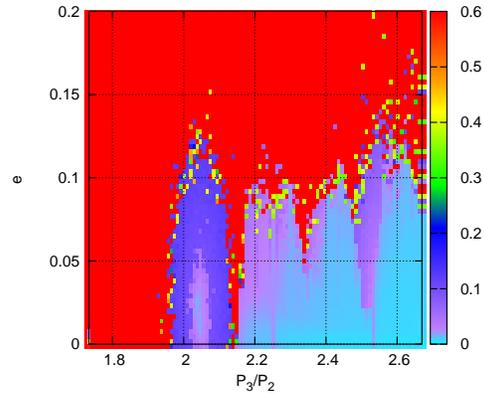}}\\
\caption{Representative plane showing general dynamical properties of Laplace resonance. The masses and eccentricities are taken from Table \ref{Table1} and the angles are: $\theta_{1}=\theta_{2}=\theta_{3}=\Delta\varpi_{31}=0$, $M_{3} = 60^{\circ})$, and $\Delta\varpi_{31}$=$180^\circ$. All the conditions survived more than $4\times10^4$ years.}
\label{fig10}
\end{figure}

Figure \ref{fig10} shows the final angle-independent representative plane for the Laplace resonance in the GJ\,876 system. We can see the real extension of the Laplace resonance because we have become independent of the angles. An island of stability can be identified in the region where the MMR 2:1 is located, where the Laplace resonance is acting as protection mechanism of the bodies. We also performed integrations of the individual 2:1 and 4:1 resonance with the two planets involved in each one. In the map of the 2:1 resonance the stability area shrinks much more than the area showed in Fig. \ref{fig10}. Thus the Laplace resonance can have important properties for stability of extrasolar planetary systems, although this doesn't mean that every system trapped in a Laplace resonance will be stabilized by these mechanism.

\section{Conclusions}
Despite the arbitrary choice of the studied system (GJ\,876), we introduced several constrains for a three-body system in a Laplace configuration. The analysis developed in this work could be easily expanded to any other system that seems to be near a three-body resonance. 

We confirmed that published solutions of the system GJ\,876 are located truly in a Laplace resonance, although their behavior is chaotic (not due to the presence of the planet with period P=1.93 days). 

Not only their Laplace angle is librating with small amplitude, also the nominal location of the systems from Table \ref{Table1} are surrounded by stable although chaotic regions as shown in Figure \ref{fig3}. We also showed that the same systems when integrated in other planes (e.g. $\Delta \varpi = 180$) are stable. Using Rivera's solution for $e$ ranging from 0.01 to 0.05 regular solutions were found. 

Through the exploration of the parameter space around best fit solutions we found that other stable Laplace solutions exist, surrounded by  unstable region which is quickly evidenced in less than 1 $\times 10^3$ years. Also preferable orientations for the libration of Laplace angle was found (see Figure \ref{fig5}).

We defined lower and upper mass limits for the outer planet ($0.02 m_{2} < m_{3} < 0.5 m_{2}$). We checked that these limits are independent on mass ratio between the two inner planets. This independence gives us enough reasons to suppose that similar stability limits can be found on other systems which are locked in this type of three body configurations. Limits in the mass ratio can be very useful for future works on problems involving three-body resonances.

The mutual inclination for the exterior planet can be from $-15^\circ$ to $-15^\circ$, being the coplanar orbits more regular solutions.  

The Laplace resonance is surrounded by chaotic motion (determined with Megno). The behavior of the variation of eccentricity seems to be a good indicator for different dynamical regimes, and we verified that for a very thin island with eccentricities $e_3$ from 0.01 to 0.04 exists a region of regular orbits. 

A representative plane for the Laplace resonance, independent of angular initial conditions was found. All angles must set to $0^\circ$ ($\theta_{1}=\theta_{2}=\theta_{3}=\Delta\varpi_{31}=0$) but $\Delta\varpi_{32}$=$180^\circ$. There exists a strict relation for the mean anomalies, i.e. $M_1=2M_2=4M_3$, although not all the values gave stable conditions. In this plane we can observed the real extension of Laplace resonance.

\section*{Acknowledgements}

Numerical simulations were made on the local computing resources (Blafis Cluster) at the University of Aveiro (Aveiro, Portugal).

\bibliographystyle{mn2e}

\bibliography{library}

\end{document}